\documentclass[aps,prl,twocolumn,nofootinbib]{revtex4-1}
\usepackage{graphicx}
\usepackage{slashed}
\newcommand{\BM}[1]{{\mbox{\boldmath{$#1$}}}}
\newcommand{\fr}[2]{{\hbox{$ #1 \over #2 $}}}

\begin{document}

\title{Hyperfine splitting in muonic hydrogen constrains new pseudoscalar interactions}

\author{W.-Y.~Keung$^{1,3}$ and D. Marfatia$^{2,3}$}
\affiliation{$^1$Department of Physics,
University of Illinois at Chicago,
 Chicago, Illinois 60607, USA}
\affiliation{$^2$Department of Physics and Astronomy,
University of Hawaii, Honolulu, Hawaii 96822, USA}
\affiliation{$^3$Kavli Institute for Theoretical Physics, University of California, Santa Barbara, California 93106, USA}

\vskip 2cm
\begin{abstract}
We constrain the possibility of a new pseudoscalar coupling between the muon and proton using a recent measurement of the 2S hyperfine splitting in muonic hydrogen.
\end{abstract}

\maketitle


Recent measurements of $2S-2P$ transition frequencies in the exotic atom constituted by a proton orbited by a muon~\cite{Pohl:2010zz,Antognini:1900ns} find the proton charge
radius to be $7\sigma$ smaller than the 2010-CODATA~\cite{codata} value obtained using ordinary hydrogen and $e-p$ 
scattering.
The 2S hyperfine splitting deduced from the same measurements shows excellent agreement with predictions~\cite{Antognini:1900ns}. The
discrepancy in the proton radius has generated a lot of interest, including the invocation of new fundamental interactions as an explanation. Here, we focus on the implications of the hyperfine splitting for new interactions between the muon and proton. Specifically, we consider the possibility of a new pseudoscalar particle that couples to the muon and proton. Such an interaction is spin and velocity dependent and has a negligible effect on the Lamb shift (which is used to extract the proton radius) in the nonrelativistic limit~\cite{Barger:2010aj}, but has a significant effect on the hyperfine splitting.

The measured value of the 2S hyperfine splitting (HFS)~\cite{Antognini:1900ns}
\begin{equation}
\Delta E_{HFS}=22.8089\pm0.0051\ \  \rm{meV}\,,
\end{equation}
is to be compared with the theoretical prediction~\cite{Antognini:2012ofa}
\begin{equation}
\Delta E^{th}_{HFS}=(22.9843\pm 0.0030)-(0.1621\pm 0.0010)r_Z+\delta E_a\,,
\end{equation}
in meV, where the Zemach radius~\cite{zemach}
\begin{equation}
r_Z=1.045\pm0.004\ \ \rm{fm}\,,
\end{equation}
 is obtained from $e-p$ scattering.{\footnote{The use of the value of $r_Z$ obtained from $e-p$ scattering is appropriate here because the correction to $r_Z$ from using the new $\mu-p$ interaction arises at loop order.} $\delta E_a$ is the contribution to HFS from the new pseudoscalar interaction. Taking the experimental and theoretical uncertainties in quadrature, the best-fit to
the experimentally measured $\Delta E_{HFS}$ and $r_Z$ occurs for $r_Z=1.045$~fm and $\delta E_a=-0.006$~meV, and
\begin{equation}
-0.018\ {\rm{meV}} \le \delta E_a \le 0.006\ \rm{meV}\ \ {\rm at}\  2\sigma.
\end{equation}

We now compute $\delta E_a$, and subject it to the above $2\sigma$ constraint.
In the nonrelativistic (NR) limit, the pseudoscalar vertex becomes
$$ J_5=\bar{u}(\BM{p}') i\gamma_5 u(\BM{p}) 
      \stackrel{\rm NR}{\longrightarrow}
       i \chi'^\dagger \fr{\BM{\sigma\cdot \BM p}}{2m}\chi 
       -i \chi'^\dagger \fr{\BM{\sigma\cdot \BM p'}}{2m}\chi  \,, $$
where $\chi$ and $\chi'$ are 2-component Pauli spinors.
The $\mu-p$ interaction in terms of the muon line (given by $\chi_\mu, \BM\sigma_\mu$) and 
the proton line (given by $\chi_p, \BM\sigma_p$) is then (see Fig.~\ref{feyn}),
$$ J_{5,\mu} J_{5,p}=
\fr{i}{2m_\mu} {\chi'}_\mu^\dagger   \BM{\sigma_\mu\cdot (p-p')}\chi_\mu  \ \
 \fr{i}{2m_p}   \chi_p'^\dagger \BM{\sigma_p\cdot (P-P')}\chi_p  \,, $$
and the NR scattering amplitude for
$\BM{p+P  \to p'+P'}$ is
$$ i{\cal M}= i f_\mu  J_{5,\mu}\ \fr{i}{q^2-m_a^2}   i f_p J_{5,p} \ , \hbox{
with } q=p-p'=P'-p  \,.$$
\begin{figure}[t]
\centering
\includegraphics[width=3.1in]{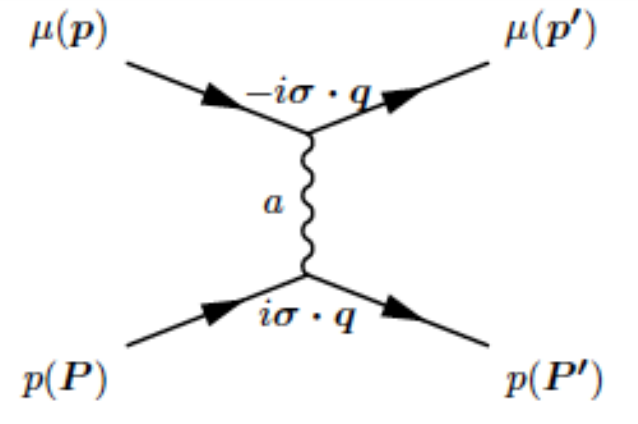}
\caption{\label{feyn}%
Pseudoscalar exchange in muonic hydrogen.}
\end{figure}

The couplings of the light
pseudoscalar $a$ of mass $m_a$ to the muon and to the proton are $f_\mu$ and $f_p$, respectively. 
Then, 
$$ {\cal M} = -\fr{f_\mu f_p}{4m_\mu m_p}
\ \  \chi_\mu'^\dagger   \BM{\sigma_\mu\cdot {q}}\chi_\mu
\ \  \chi_p'^\dagger \BM{\sigma_p\cdot  {q}}\chi_p 
\ \ \fr{1}{\BM{q}^2+m_a^2}   \ $$
$$ \qquad =
-\fr{f_\mu f_p}{4m_\mu m_p}  \frac{1}{3} \BM{q}^2 
\ \  \chi_\mu'^\dagger   \BM{\sigma_\mu}\chi_\mu \ \cdot   
\ \  \chi_p'^\dagger \BM{\sigma_p}\chi_p                              
\ \ \fr{1}{\BM{q}^2+m_a^2}   \ ,$$
with the relative angle averaged for the $s$ wave.
The effective Hamiltonian is
$$ \delta H_a=\frac{1}{3}   \frac{f_\mu f_p}{4m_\mu m_p}\
                   \left[\delta^3(\BM{r}) -\frac{m_a^2 e^{-m_a r}}{4\pi r}
                  \right]
                   \BM{\sigma_\mu \cdot \sigma_p}   \,, $$
so that
$$ \delta E_a 
  = \frac{f_\mu f_p}{3 m_\mu m_p}  \left[
   |\psi(0)|^2  - m_a^2 \int|\psi(\BM{r})|^2 \frac{e^{-m_a r}}{4\pi r} 
d^3\BM{r}     \right]
\,, $$
where $\psi$ is the wave function of the $2S$ state:
$$ \psi(\BM{r})=\fr1{2\sqrt{2\pi a_B^3}} (1-\fr{r}{2 a_B})e^{-\fr{r}{2a_B}} \,. $$
Here, $a_B=\fr{1}{\alpha m_r} $ is the Bohr radius for muonic hydrogen with $m_r=m_\mu m_p/(m_\mu+m_p)$, the
reduced mass of the system.
On convolving, we obtain
\begin{equation}
 \delta E_a =\frac{f_\mu f_p \alpha^3 m_r^3} {3 m_\mu m_p}
\frac{1}{8\pi}  F \left(\frac{m_a}{m_r}\right)\,,
\end{equation}
where
\begin{equation}
F(x)=1- x^2 \frac{\alpha^2+2x^2}{2(\alpha+x)^4} \,.\nonumber
\end{equation}
It is important to distinguish between $m_r$ and $m_\mu$ in the equations above.
The $m_r$ dependence comes from the Bohr radius $a_B$, and $m_\mu$ 
from the NR reduction.
The function $F(x)$ interpolates between 1 and 0 for $x=0$ and $x\to\infty$ which is consistent with  decoupling behavior.

\begin{figure}[t]
\centering
\includegraphics[width=3.3in]{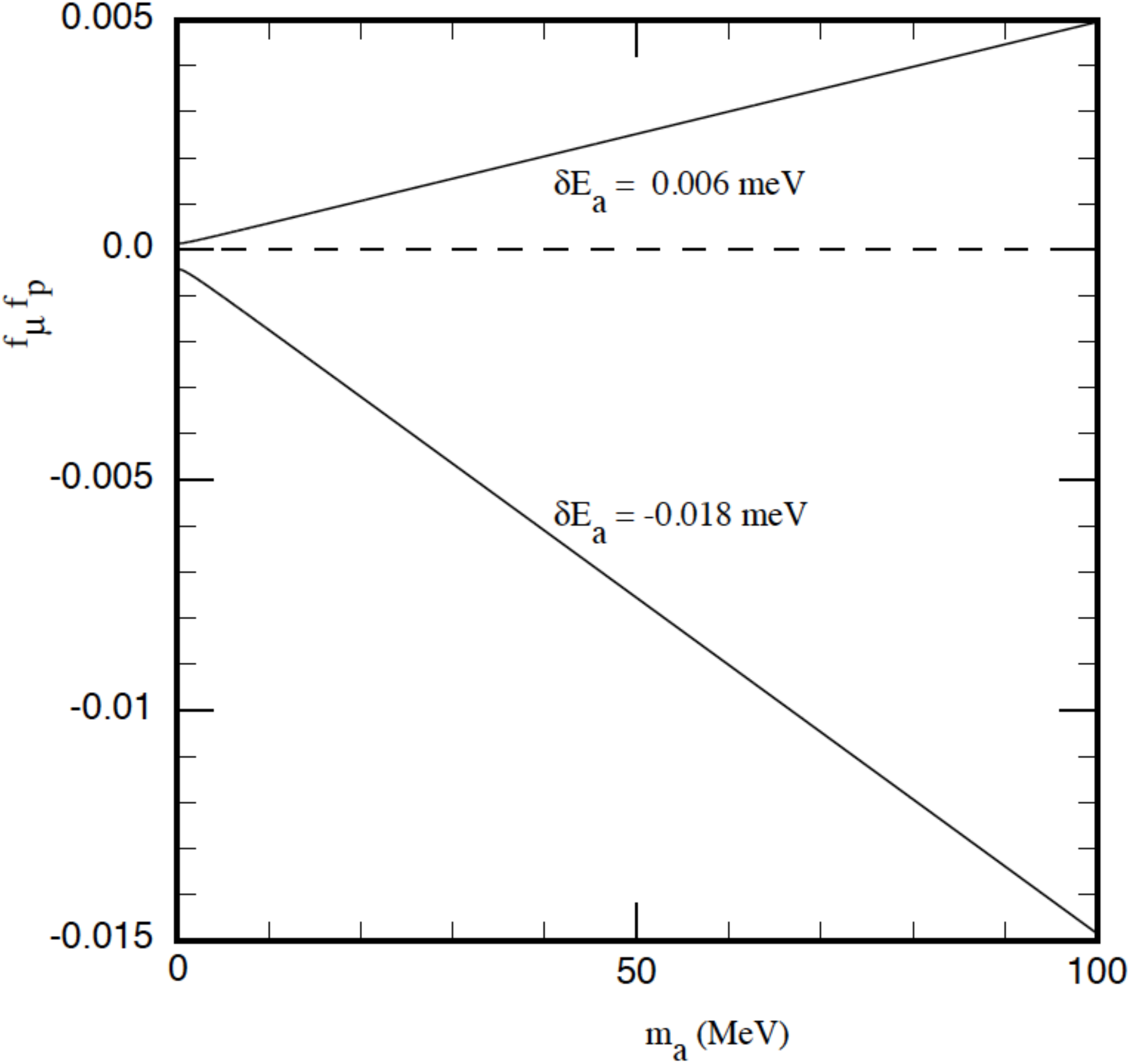}
\caption{\label{constraint}%
The values of $f_\mu f_p$ allowed at $2\sigma$ lie between the solid curves.}
\end{figure}

In Fig.~\ref{constraint}, we show the $2\sigma$ allowed values of $f_\mu f_p$ as a function of $m_a$. The region between the solid curves is allowed. We restrict $m_a \le 100$~MeV so that $f_\mu f_p$ remains comfortably in the
perturbative regime.

Note that in the potential model  of the proton with nonrelativistic quarks,
the proton pseudoscalar coupling $f_p$ arises from the pseudoscalar
couplings $f_u,f_d$ of the up and down quarks, which are of the same order of magnitude. In this simplified picture, we have
$f_p=\fr43 f_d-\fr13 f_u$ (as for the magnetic moments).
 If $f_u=f_d$, we have the simple result, $f_p=f_u=f_d$.

In principle, the anomalous magnetic moment of the muon places a stringent independent constraint on $f_\mu$ since
for the $m_a$ of interest, pseudoscalar couplings yield a negative contribution to $a_\mu$~\cite{leveille,w-y},\footnote{In Eq.~(11) of Ref.~\cite{leveille},
$C_P^2$ should be replaced by $|C_P|^2$, since $C_P$, as defined in Eq.~(9) therein, is complex.}  while the measured value is {\it higher} than the standard model expectation: $\Delta a_\mu=a_\mu^{exp}-a_\mu^{th}=(29\pm 9)\times 10^{-10}$~\cite{g2}.
However, the scalar sector may be more intricate than envisioned here, and may offer a fine-tuned (and perhaps unnatural) cancellation of the pseudoscalar contribution.

For the sake of comparison,  the QED contribution at leading order is
$$ \delta H_{\rm QED}= \frac{e^2}{6} \frac{ g_\mu g_p }{4m_\mu
  m_p}\ \delta^3(\BM{r}) \ 
{\BM{\sigma_\mu \cdot \sigma_p}} \ . $$
Here, $g_\mu$ ($\approx 2$), and $g_p$ ($\approx 5.5857$) are the gyromagnetic ratios for the
muon and proton. Correspondingly,
\begin{eqnarray}
\delta E_{\rm QED}
&=&\frac{\alpha^4 m_r^{3}} {12 m_\mu m_p} g_\mu g_p\,.\nonumber
\end{eqnarray}
 The above QED result, though simple, 
represents the first three significant digits of 
the dedicated theoretical 
calculation,  and is consistent with the recent measurement of Ref.~\cite{Antognini:1900ns}. 

The ratio of the pseudoscalar contribution to the leading QED contribution is
\begin{eqnarray}
 \frac{\delta E_a}{\delta E_{\rm{QED}}}
 &=&\frac{2}{4\pi\alpha}
    \frac{f_\mu f_p}{g_\mu g_p}\
        {\Large F}\left(\frac{m_a}{m_r}\right)\nonumber\\
  &=& \frac{2}{4\pi\alpha}  
     \frac{f_\mu f_p}{g_\mu g_p}  \left[
   1 -    \frac{m_a^2}{m_r^2}
    \frac{\alpha^2+2(m_a/m_r)^2}{2(\alpha+m_a/m_r)^4} \right]\,. \nonumber
\end{eqnarray}

In sum, the 2S hyperfine splitting in muonic hydrogen constrains the product of the pseudoscalar couplings of the muon and proton $f_\mu f_p$ to lie in the $2\sigma$ ranges $[-0.00040,0.00013]$, $[-0.00173,0.00058]$ and $[-0.015,0.005]$  for $m_a= 0$, 10~MeV and 100~MeV, respectively. As the pseudoscalar mass is further increased, the constraint is weakened. The couplings have no impact on the discrepant measurements of the proton radius. 

For $m_a < 100$~MeV, no direct limits on $f_\mu f_p$  exist from colliders, although limits for higher $m_a$ were obtained by the CMS experiment using the dimuon channel in $pp$ collisions~\cite{cms}. The CMS upper limits on the cross section times
branching fraction, $\sigma \cdot B(pp \to a \to \mu^+\mu^-)$, directly constrain $f_\mu f_p$ in the mass ranges, $5.5-8.8$~GeV and $11.5-14$~GeV. 
CLEO's nonobservation of the decay \mbox{$J/\psi \to \gamma a$}, with $a$ invisible~\cite{cleo} gives the 90\%~C.L. constraint $|f_p|<0.029$ (assuming the $J/\psi - a$ coupling to be $f_p$) for $m_a<100$~MeV~\cite{Barger:2010aj}. 



{\it Acknowledgments.}
This work was supported by the DOE under
Grant Nos. DE-SC0010504 and DE-FG-02-12ER41811, and by the Kavli Institute for Theoretical Physics under NSF Grant No. PHY11-25915.

\end{document}